\documentclass[floats,aps,prl,psfig,graphicx,twocolumn]{revtex4}
\usepackage{graphicx}
\usepackage{amsmath}

\begin{document}

\title{Ab Initio Simulations of Hot, Dense Methane During Shock Experiments}

\author{Benjamin L. Sherman$^1$, Hugh F. Wilson$^2$, Dayanthie Weeraratne$^1$, and Burkhard Militzer$^{2,3}$} 

\affiliation{$^1$ Department of Geological Sciences, California State University Northridge, Northridge, CA 91330, USA}
\affiliation{$^2$ Department of Earth and Planetary Science, University of California, Berkeley, CA 94720, USA}
\affiliation{$^3$ Department of Astronomy, University of California, Berkeley, CA 94720, USA}

\author{Using density functional theory molecular dynamics
  simulations, we predict shock Hugoniot curves of precompressed
  methane up to 75$\,$000 K for initial densities ranging from 0.35 to
  0.70 g$\,$cm$^{-3}$. At 4000~K, we observe the transformation into a
  metallic, polymeric state consisting of long hydrocarbon chains.
  These chains persist when the sample is quenched to 300$\,$K,
  leading to an increase in shock compression. At 6000$\,$K, the
  sample transforms into a plasma composed of many, short-lived
  chemical species. We conclude by discussing implications for the
  interiors of Uranus and Neptune and analyzing the possibility of
  creating a superionic state of methane in high pressure
  experiments.}

\maketitle

Methane is one of the most abundant chemical species in the universe,
with the vast majority of our solar system's share being locked away
in the interiors of the giant planets~\cite{hubbard-81}. While recent
work has suggested that icy components of the cores of gas giants such
as Jupiter and Saturn are likely to be dissolved into the surrounding
hydrogen~\cite{WilsonMilitzer2012,WilsonMilitzer2012b}, the ice giants
Uranus and Neptune likely contain a significant amount of methane at
pressures ranging from 20 to 80 GPa and 2000 to 8000~K.  Extrasolar
planets in the ice giant mass regime are now known to be
extremely common throughout the universe~\cite{borucki}. This
motivates our need for a greater understanding of the behaviour of
this important material in this specific range of pressures and
temperatures.

Dynamic shock experiments~\cite{Hicks2006,Knudson2009} combined with
static precompression~\cite{KananiLee2006,MH08,jeanloz2007} have the
potential to probe the behavior of methane at pressures and
temperatures in ice giant interiors but these experiments are
challenging and their interpretation often not
straightforward~\cite{Spaulding2012,Militzer2012}. There exists a
long tradition of first-principles simulation predicting the outcome
of past~\cite{MC00,Mi01,Knudson2009,Mi09} and future~\cite{Mi06} shock
experiments and providing an interpretation on the atomistic level for
the thermodynamic changes that were observed~\cite{Mi03}. Here we use
density functional molecular dynamics (DFT-MD) simulations to
characterize the properties of dense, hot methane and identify a
molecular, a polymeric, and a plasma state. From the computed equation
of state, we predict the shock Hugoniot curves for different initial
densities in order to guide future shock wave experiments with
precompressed samples.

The high pressure, high temperature behavior of methane has previously
been probed experimentally and theoretically across a range of
conditions, revealing a complex chemical behavior characterized by the
conversion of methane into higher hydrocarbons and eventually diamond
as pressure is increased. Laser heated diamond anvil cell (LHDAC)
studies by Benedetti~\emph{et al.}~\cite{benedetti} showed the
presence of polymeric hydrocarbons and diamond at pressures between 10
and 50 GPa and temperatures of about 2000 to 3000~K. More recently, a
LHDAC experiment by Hirai~\emph{et al.}~\cite{hirai} found evidence of
the formation of carbon-carbon bonds occurring above 1100 K and 10
GPa, including the formation of diamond above 3000~K. Gas gun shock
wave experiments by Nellis~\emph{ et al.}~\cite{nellis} achieved
pressures of 20 to 60 GPa and produced results that were interpreted
as the formation of carbon nanoparticles. Radousky~\emph{et
  al.}~\cite{radousky} previously carried out shock experiments at
lower pressures that did not reach the dissociation
regime. 

Ancilotto~\emph{et al.}~\cite{ancilotto} performed DFT-MD
simulations that followed the predicted isentropes of the intermediate
layers of Uranus and Neptune. At 100 GPa and 4000 K methane was found
to dissociate into a mixture of hydrocarbons. At 300 GPa and 5000 K
it became apparent that carbon-carbon bonds were favored with
increasing pressure. Recent DFT-MD simulations with free energies
derived from the velocity autocorrelation function by Spanu \emph{et
  al.}~\cite{spanu} predicted that higher hydrocarbons than methane
are energetically favored at temperature-pressure conditions between
1000 and 2000 K and at 4 GPa and above. Static zero-kelvin
calculations by Gao~\emph{et al.}~\cite{gao} predicted methane to
decompose into ethane at 95 GPa, butane at 158 GPa and diamond at 287
GPa. Goldman~\emph{et al.}~\cite{goldman} computed shock Hugoniot
curves for the molecular regime of methane up to 50 GPa. Better
agreement with experiments was obtained by approximately including
quantum corrections to the classical motion. The vibrational spectrum
for each atomic species was derived from DFT-MD simulations and the
internal energy was corrected by assuming each atom has six harmonic
degrees of freedom. This approach is exact for a harmonic solid but
its validity remains to be more carefully evaluated for molecular fluids.

Laser and magnetically driven shock
experiments~\cite{Hicks2006,Knudson2009} now allow multi-megabar
pressures to be obtained routinely, however the lack of independent
control of the pressure and temperature variables makes the direct
study of planetary interior conditions
difficult~\cite{MH08,jeanloz2007}. For any given starting material,
the pressure and temperature conditions of the shocked material are
uniquely determined by the shock velocity. Conservation of mass, momentum and
energy across the shock front lead to the Hugoniot relation~\cite{zeldovich},
\begin{equation}
H = (E - E_0) + \frac{1}{2}(P + P_0)(V - V_0) = 0,
\label{eq1}
\end{equation}
relating the final values of the internal energy, $E$, pressure, $P$,
and volume, $V$, to the corresponding values $E_0, P_0$, and $V_0$ for
the initial state of the material. While shocks launched into
materials at ambient conditions allow extremely high pressures to be
reached, a large part comes from the thermal pressure and the density
is often only about 4 times higher than the initial value~\cite{Mi06}.
The resulting shock temperatures are significantly higher than those
along the planetary isentrope. Static precompression of the sample,
however, shifts the Hugoniot curves to higher densities allowing
higher pressures to be explored for given temperature. Precompressed
shock experiments on methane are thus highly desirable as a means of
exploring the pressure-temperature conditions of interest to planetary
science.

In this article, we present theoretically computed methane shock
Hugoniot curves corresponding to initial densities ranging from 0.35
to 0.70 g$\,$cm$^{-3}$ and temperatures up to 75$\,$000 K. We use DFT-MD
simulations to determine equations of state of methane at relevant
temperature-pressure conditions, compute the equation of state, and
analyze the atomistic and electronic structure of the hot, dense fluid
in order to make predictions for future shock measurements using methane.

\section{Computational methods}

All DFT-MD simulations in this article were performed with the Vienna
Ab Initio Simulation Package (VASP)~\cite{vasp}. Wavefunctions were
expanded in a plane wave basis with an energy cutoff of 900~eV, with
core electrons represented by pseudopotentials of the {projector
  augmented wave} type~\cite{paw}. Exchange-correlation effects were
approximated with the functional of Perdew, Burke, and
Ernzerhof~\cite{pbe}. Each simulation cell consisted of 27 methane
molecules in a cubic cell with temperature controlled by a Nose-Hoover
thermostat. The Brillouin zone was sampled with $\Gamma$ point only in
order to invest the available computer time into a simulation cell
with more atoms. Electronic excitations were taken into account using
Fermi-Dirac smearing~\cite{mermin}.
	
\begin{figure}[htbl]
\includegraphics[width=0.47\textwidth]{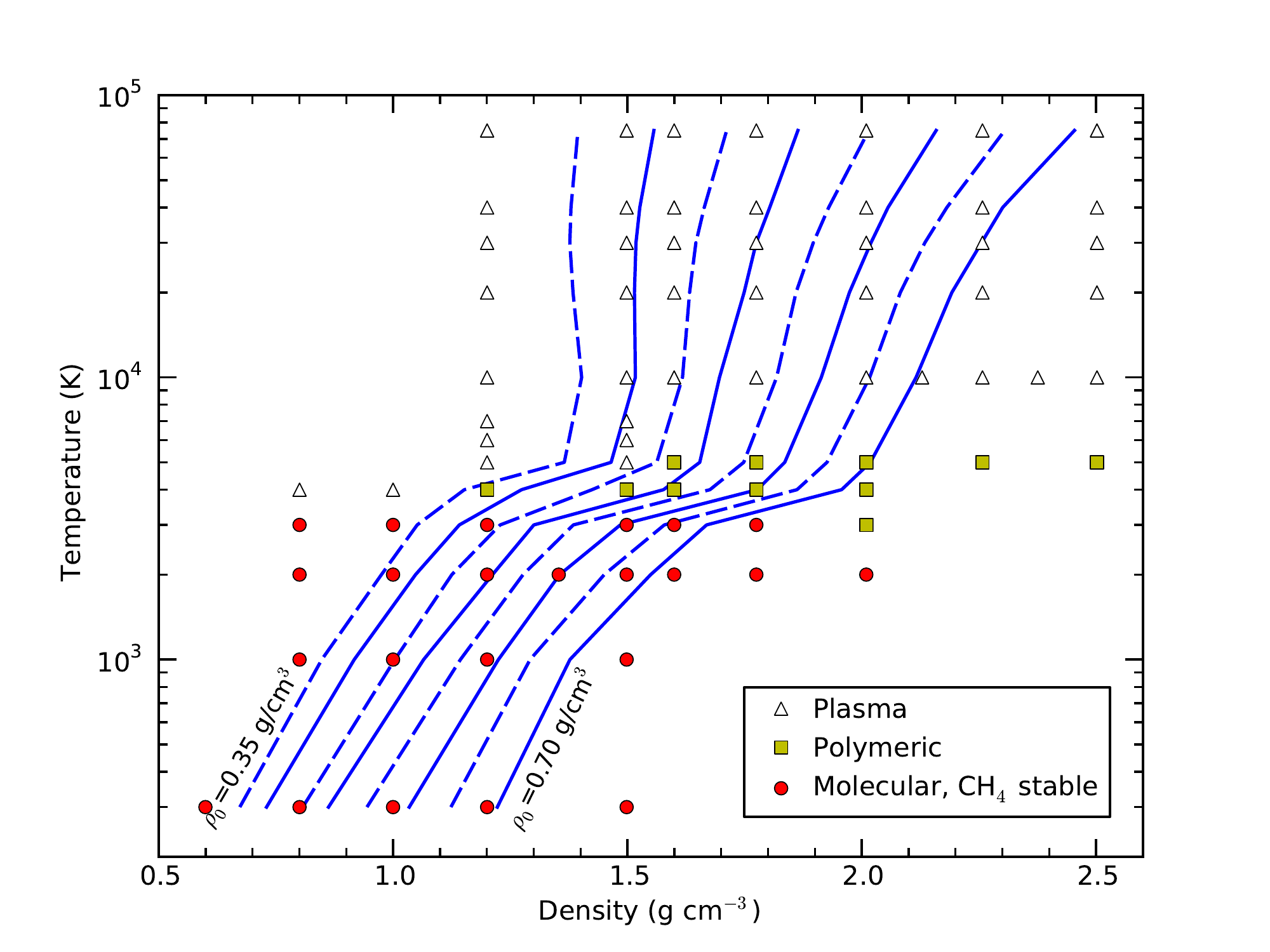}
\caption{Methane shock Hugioniot curves showing temperature as a
  function of density for different initial densities,
  $\rho_0=\{0.35,0.40,0.45,0.50,0.55,0.60,0.65,0.70\}$. The symbols
  show the density and temperature conditions of our DFT-MD
  simulations that either stayed molecular, or reached a polymeric or
  plasma state.}
\label{figure1}
\end{figure}

Simulations were undertaken at 79 different temperature-density
conditions as shown in Fig.~\ref{figure1}, chosen in order to
encompass the Hugoniot curves for methane at initial densities ranging
from 0.35$-$0.70 g$\,$cm$^{-3}$. Temperatures ranged from 300 to
75$\,$000~K and pressures from 3 to 1100 GPa. Each simulation lasted
between 1 and 2~ps, and in some cases up to 8 ps. Almost all
simulations used an MD time step of 0.2~fs. Initially, we performed a
few simulations with a time step of 0.4~fs at lower temperatures but
reduced the timestep as we explore higher temperature and densities.
Our tests showed that the computed thermodynamic functions at lower
temperatures agreed within error bars with the results obtained with
the shorter time step.

\section{Results}

The pressure and internal energy computed from DFT-MD simulations at
various densities and temperatures are given in Tab.~\ref{tab1} in the
online supplemental information. Any initial transient behavior was
removed from the trajectories before the thermodynamic averages were
computed. Hugoniot curves were calculated by linear interpolation of
the Hugoniot function, $H$, in Eq.~(\ref{eq1}) as a function of
temperature and pressure. Figure~\ref{figure1} shows temperature as a
function of density along the Hugoniot curves generated for each of
the eight initial densities, $\rho_0$. Several features are
immediately apparent. As might be expected, precompression allows
higher densities to be reached at any given temperature. Each Hugoniot
curve can be divided into three segments. Up to 3000~K, one finds the
temperature to increase linearly with density. In our simulations, the
methane molecules remained intact. At a temperature of approximately
4000--5000~K, a plateau is reached where density increases rapidly
without a corresponding increase in temperature. Our simulations show
the system entering into a polymeric regime where the methane
molecules spontaneously dissociate to form long hydrocarbon chains
that dissociate and reform rapidly. We will demonstrate this regime to
be metallic, which implies a high electrical conductivity and
reflectivity should be detected during a shock experiments.

\begin{figure}[htbl]
\includegraphics[width=0.47\textwidth]{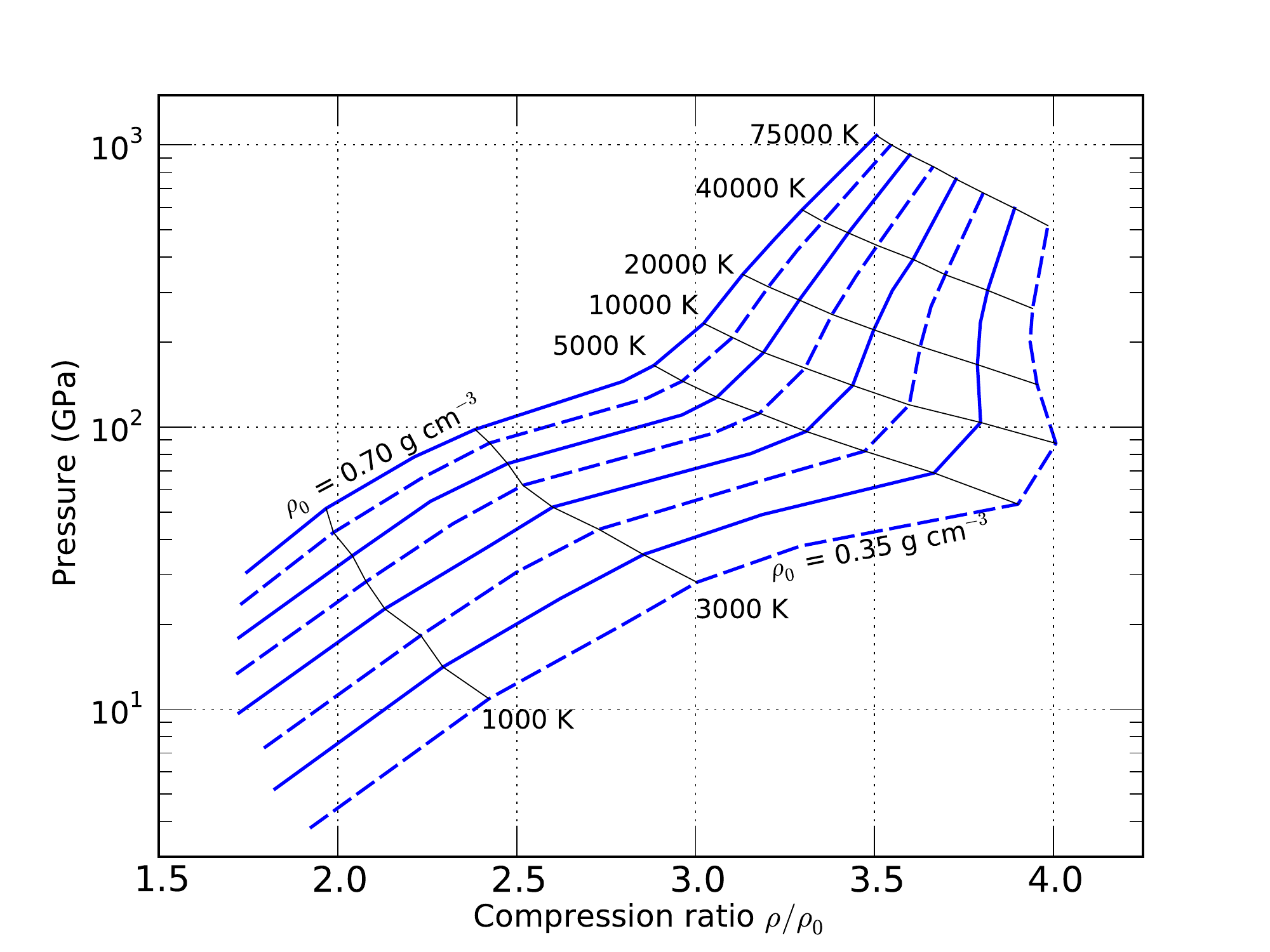}
\caption{The pressure as function of the compression ratio,
  $\rho/\rho_0$, along the shock Hugoniot curves from
  Fig.~\ref{figure1} for different initial densities, $\rho_0$.}
\label{figure3old}
\end{figure} 

In the third regime, the temperature on the Hugoniot curve again
increases rapidly as the system assumes a plasma state where many
ionic species exist for a very short time. The slope of the curve
significantly depends on the initial density. For a small degree of
precompressions, density on the Hugoniot curve does not 
increase in the plasma regime because the sample has already been
4-fold compressed. For a high degree of precompression, density of the
Hugoniot curve keeps increasing in the plasma regime because the
theoretical high temperature limit of 4-fold compression has not yet
been reached. This is confirmed in Fig.~\ref{figure3old} where we
plotted the shock pressure as a function of compression ratio,
$\rho/\rho_0$. The highest compression ratio of 4.0 is obtained for
10$\,$000$\,$K and the lowest initial density. In general, the
compression ratio is controlled by the excitation of internal degrees
of freedom that increase the compression and interaction effects that
 act to reduce it~\cite{Mi06}. The strength of the interactions increases with
density, which explains why the compression ratio decreases when the
sample is precompressed.


\begin{figure}[htbl]
\includegraphics[width=0.47\textwidth]{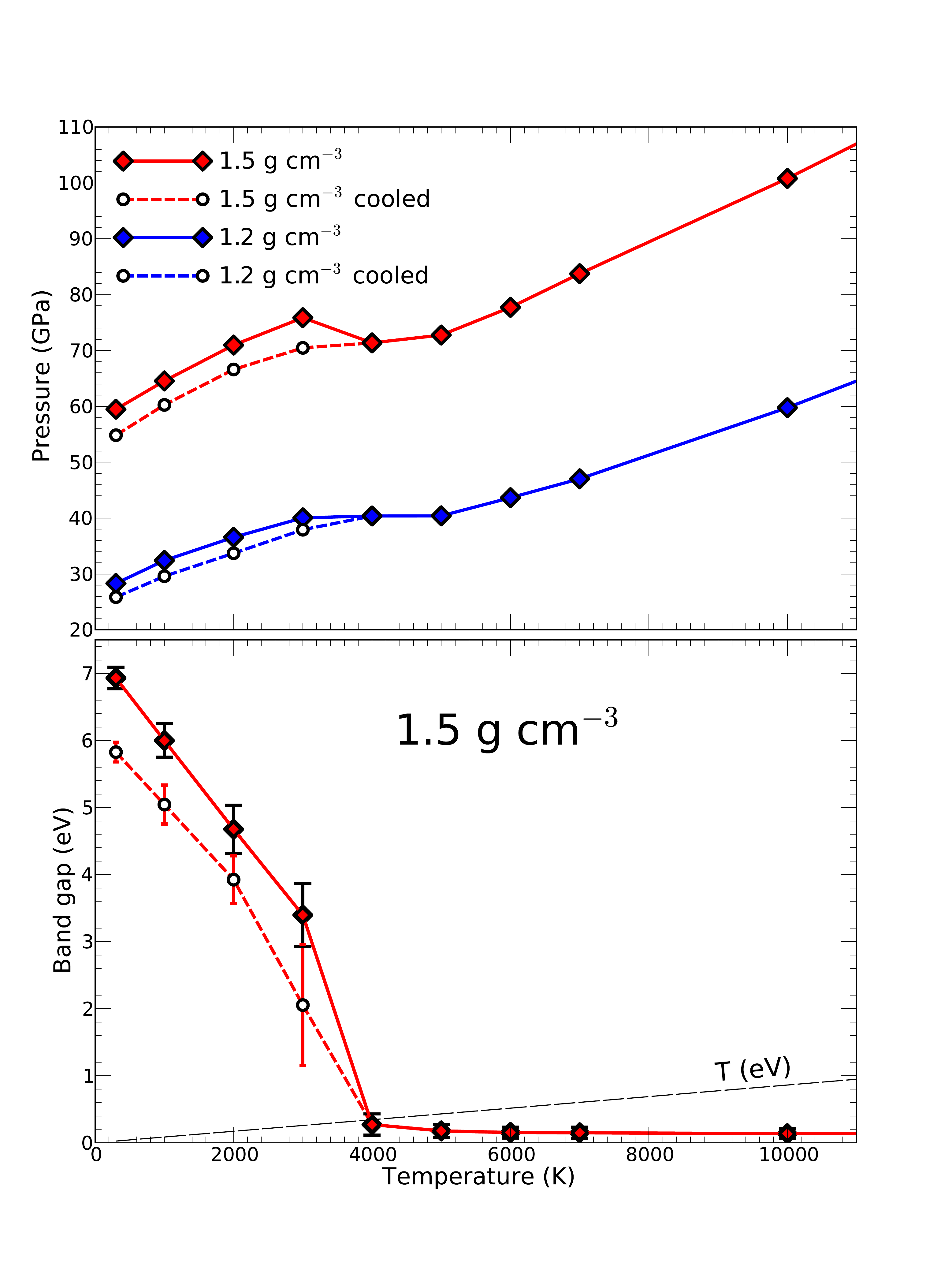}
\caption{The upper panel shows the pressure-temperature relation of
  CH$_4$ DFT-MD simulations that have been heated (filled symbols) and
  of samples that have been cooled (open symbols) from a polymeric
  state at 4000~K. The lower panel shows the corresponding electronic
  gap bands and its variance during the MD simulations. For $T \ge
  4000\,$K, the average band gap that is less than the temperature
  (dashed line), which implies that the polymeric and the plasma state
  are good electrical conductors.}
\label{TP5}
\end{figure}

We now focus on characterizing the polymeric state that gives rise to
the predicted increase in shock compression at 4000~K. Fig.~\ref{TP5}
shows a linear increase of pressure with temperature as long as the
methane molecules remain intact. When the molecules dissociate at
4000~K, the pressure reaches a plateau at 40 GPa for
1.2~g$\,$cm$^{-3}$ before it rises gradually above 5000~K when the
system turns into a plasma. For a higher density of
1.5~g$\,$cm$^{-3}$, one observes a significant drop in pressure when
molecules disassociate that resembles the behavior of molecular
hydrogen.  DFT-MD simulations of hydrogen also show that the
dissociation of molecules leads to a region with $\partial P /
\partial T|_V<0$ at high density because hydrogen atoms can also be
packed more efficiently than molecules~\cite{MHVTB,Morales2010}.

When the simulations are cooled from 4000~K, the system remained in a
polymeric state that exhibits a lower pressure than the original
simulations with methane. This confirms the free energy calculations
in Ref.~\cite{spanu} that showed the polymeric state to be
thermodynamically more stable.

In Fig.~\ref{TP5}, we also plot the electronic band gap averaged over
the trajectory at different temperatures. From 300 to 3000~K, the
system remains in an insulating state with a gap of 3~eV or larger.
When the system transforms into the polymeric state at 4000~K, the gap
drops to a value close to zero and remains small as the temperature is
increased further and the system transforms into a plasma. One may
conclude that this insulator-to-metal transition is driven by the
disorder that the dissociation introduced into the fluid and resembles
recent findings for dense helium~\cite{StixrudeJeanloz08,Mi09} where
the motion of the nuclei also led to a band gap closure at
unexpectedly low pressures.

If the band gap energy is comparable to $k_BT$, the system becomes a
good electrical conductor and its optical reflectivity increases.
Since reflectivity measurements under shock
conditions~\cite{celliers2010} have now become well established, the
transformation into a polymeric state at 4000~K can directly be
detected with experiments. However, if this transformation already
occurs at lower temperature, the polymeric transition will be more
difficult to detect with optical measurements alone because
Fig.~\ref{TP5} shows that a band gap opens back up as the polymeric
state is cooled.

\begin{figure}[htbl]
\includegraphics[width=0.47\textwidth]{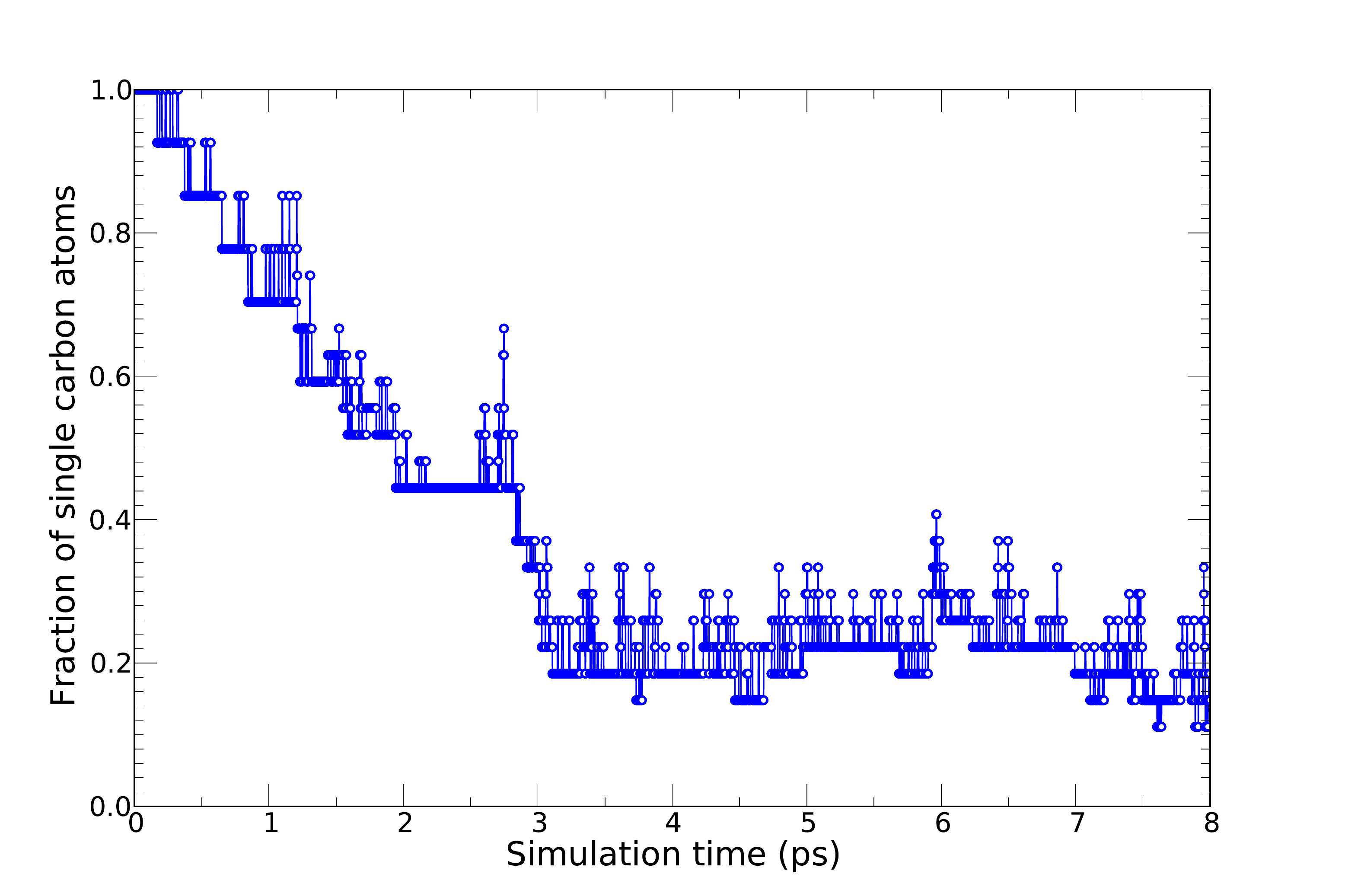}
\caption{Evolution of fraction of single carbon atoms during the MD
  simulation at 4000$\,$K and 1.2$\,$g$\,$cm$^{-3}$. The decrease
  measures how many methane molecules polymerize to form long carbon
  chains, which is illustrated in Fig.~\ref{combined}.}
\label{CH276_C1a}
\end{figure}

We now study the transformation into the polymeric state in more
detail by analyzing the clusters in the fluid. Most simply, one can
detect different polymerization reactions by analyzing the carbon
atoms only. Using a distance cut-off of 1.8 \AA, we determined how
many C$_n$ clusters were present in each configuration along different
MD trajectories. Methane molecules are classified as C$_1$ clusters in
this approach. Figure~\ref{CH276_C1a} shows a drop in the fraction of
C$_1$ clusters as the system at 4000~K and 1.2~g$\,$cm$^{-3}$
transforms into a polymeric state over the course of approximately of
3~ps.

\begin{figure}[htbl]
\includegraphics[width=0.35\textwidth]{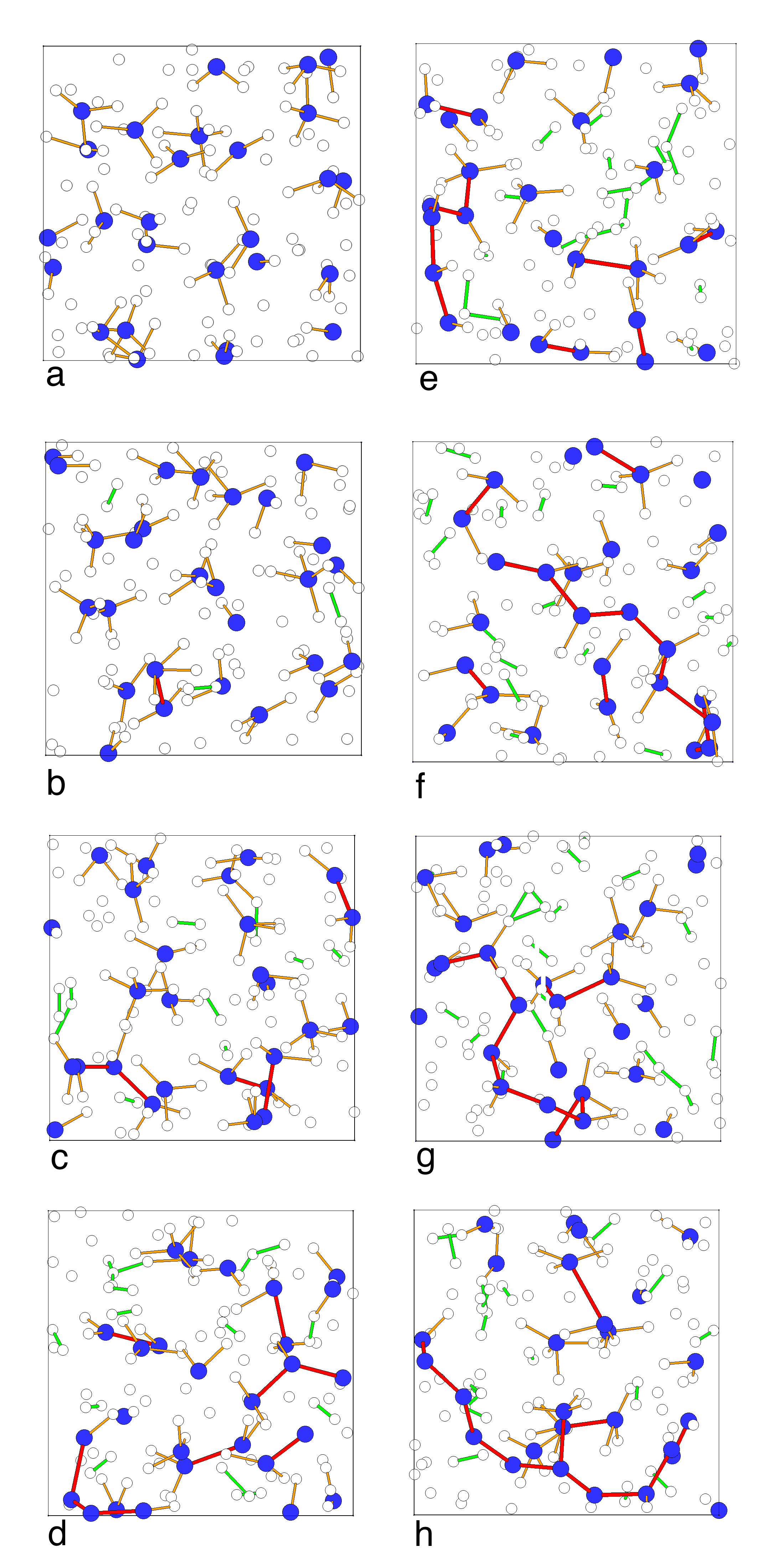}
\caption{(color online) Series of snapshots from MD simulations at
  4000$\,$K and 1.2$\,$g$\,$cm$^{-3}$. The large and small spheres
  depict the carbon and hydrogen atoms, respectively. The C-C, C-H,
  and H-H bonds are illustrated by dark thick lines, thin lines, and
  thick light lines, respectively. Panel (a) depicts the initial
  configuration with 27 CH$_4$ molecules. (b) shows the formation of
  the first ethane molecule. Some H$_2$ molecules have also formed.
  (c) and (d) show the first C$_3$ and C$_4$ chains that formed. (e-g)
  illustrate linear C$_6$, C$_7$, and C$_{10}$ chains while (h) shows a 12
  atoms chain with one branching point.}
\label{combined}
\end{figure}

In Fig.~\ref{combined}, we show a series of snap shots from the same
DFT-MD simulation in order to illustrate how hydrocarbons chain of
increasing length are formed. As soon as the first C$_2$ cluster
occurs, we also find molecular hydrogen to be present. This resembles
typical dehydrogenation reactions such as $n$CH$_4 \to$
C$_n$H$_{2n+2}+(n-1)$H$_2$ that were also analyzed in~\cite{spanu}.
This implies that molecular hydrogen may be expelled as a methane rich system
transforms into a polymeric state.

\begin{figure}[htbl]
\includegraphics[width=0.47\textwidth]{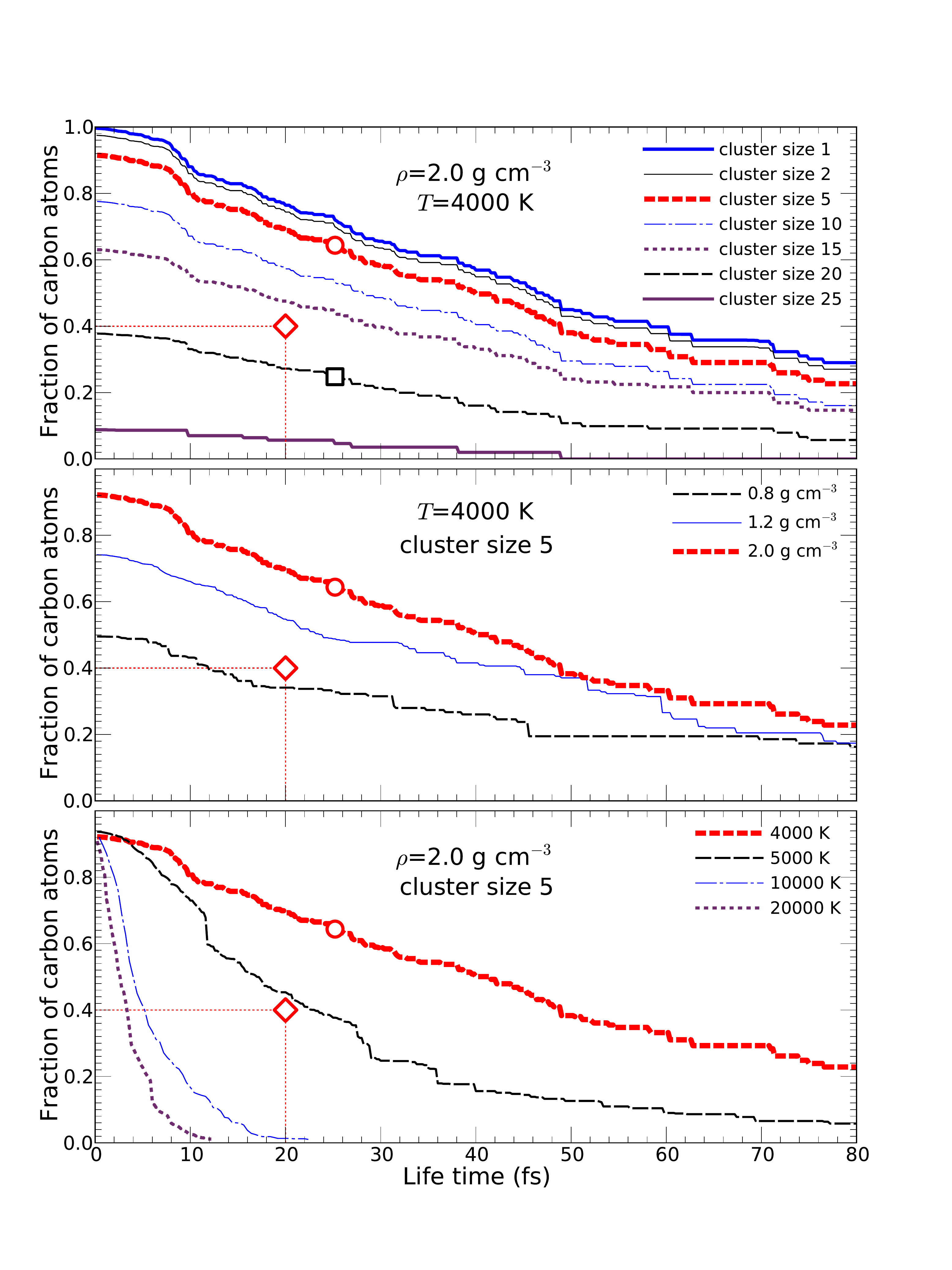}
\caption{Cumulative lifetime distribution of carbon clusters of
  different sizes at various densities and temperatures. The square in
  the upper panel shows that, on average, 26\% of the carbon atoms are
  bound in carbon clusters with 20 atoms or more that exit for at
  least 25 fs. Similarly, the open circle illustrates that 66\% of the
  atoms are bound in clusters with 5 atoms or more. Simulations were
  characterized as polymeric if at least 40\% of the atoms were bound
  in clusters with 5 atoms or more that existed for at least 20 fs (diamond symbol).}
\label{cluster_figure4}
\end{figure}

In the polymeric regime, the system is primarily composed of different
types of hydrocarbon chains but each chain exists only for a limited
time before it decomposes and new chains are formed. Every chemical
bond has a finite lifetime, which makes it similar to a plasma, but
many chains are present in every given snap shot, which may be a
unique feature of carbon rich fluids and makes the distinction from a
typical plasma state worthwhile. 

We computed the lifetime of each C$_n$ cluster in the DFT-MD
simulation, neglecting any information that may be stored in the
position of the hydrogen atoms. In Fig.~\ref{cluster_figure4} we plot
the fraction of carbon atoms that were bound in clusters of different
size and lifetimes. The plot is cumulative in two respects. The curve
for cluster size $n$ presents the average fraction of carbon atoms in
clusters with at least $n$ atoms with a lifetime at least as long as
the lifetime specified on the ordinate. By definition, the C$_1(t)$
curve approaches 1 for small $t$. It is surprising, however, that, in
simulations at 2.0~g$\,$cm$^{-3}$ and 4000~K, 26\% of all carbon atoms
were bound in clusters with 20 atoms or more that existed for at least
25~fs.

\begin{figure}[htbl]
\includegraphics[width=0.47\textwidth]{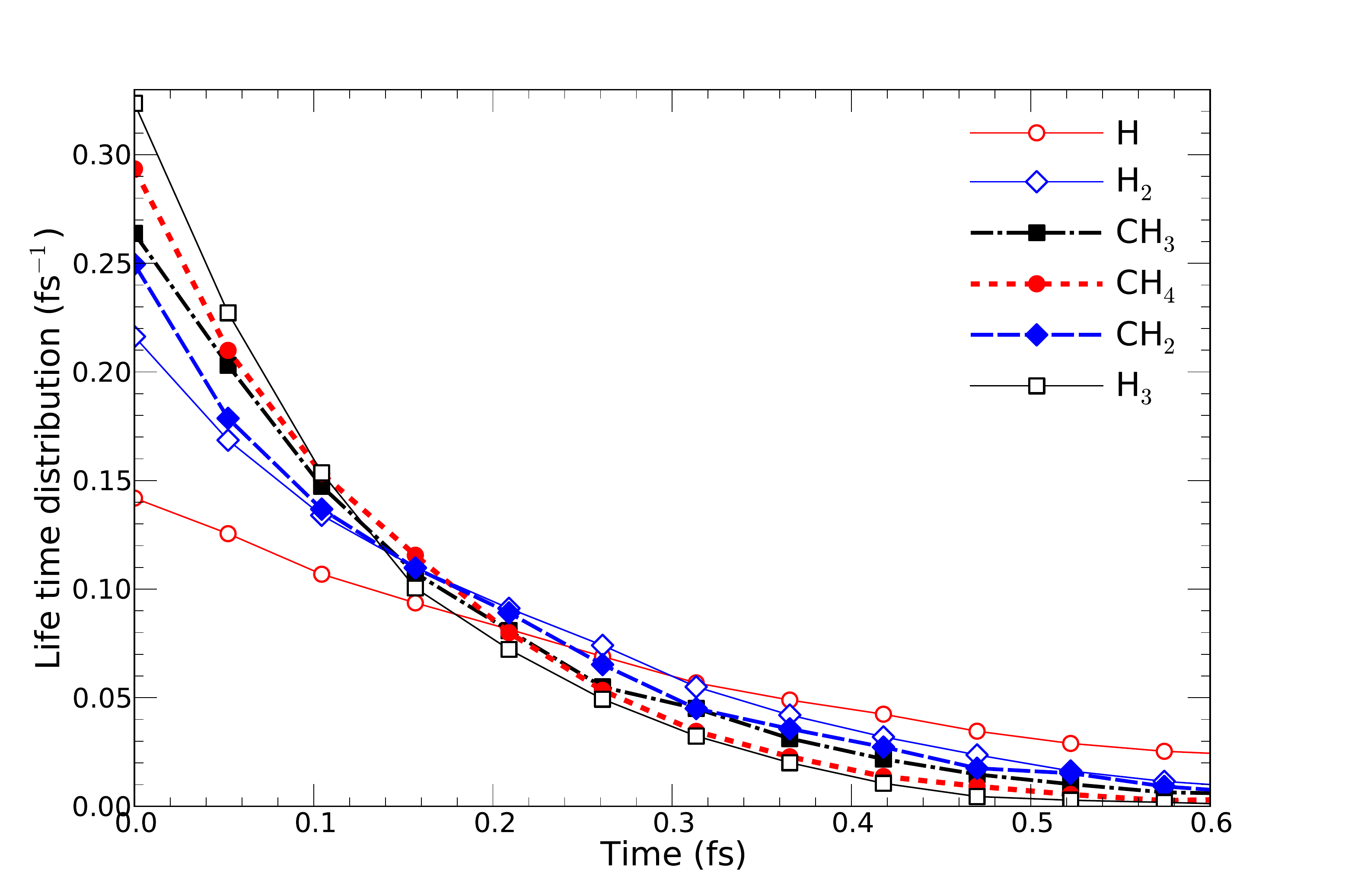}
\caption{Lifetime distribution of plasma species at 20$\,$000$\,$K and 1.5$\,$g$\,$cm$^{-3}$. }
\label{plasma}
\end{figure}

We characterized a simulation as polymeric if, on average, at least
40\% of the carbon atoms were at any given time bound in clusters with 5 atoms or more
with lifetimes of at least 20 fs. All simulations with cluster size 5
curves that fall above the diamond symbol in
Fig.~\ref{cluster_figure4} are then considered to be polymeric. The
minimum cluster size of 5 atoms distinguishes the polymeric state from
simulations with stable methane molecules while the lifetime
constraint distinguishes it from a plasma where one finds many small
ionic species that exist for less than 1~fs. Figure~\ref{plasma} shows a
typical lifetime distribution of the most common ionic species that
we obtained from a cluster analysis that includes C and H atoms.

\begin{figure}[htbl]
\includegraphics[width=0.47\textwidth]{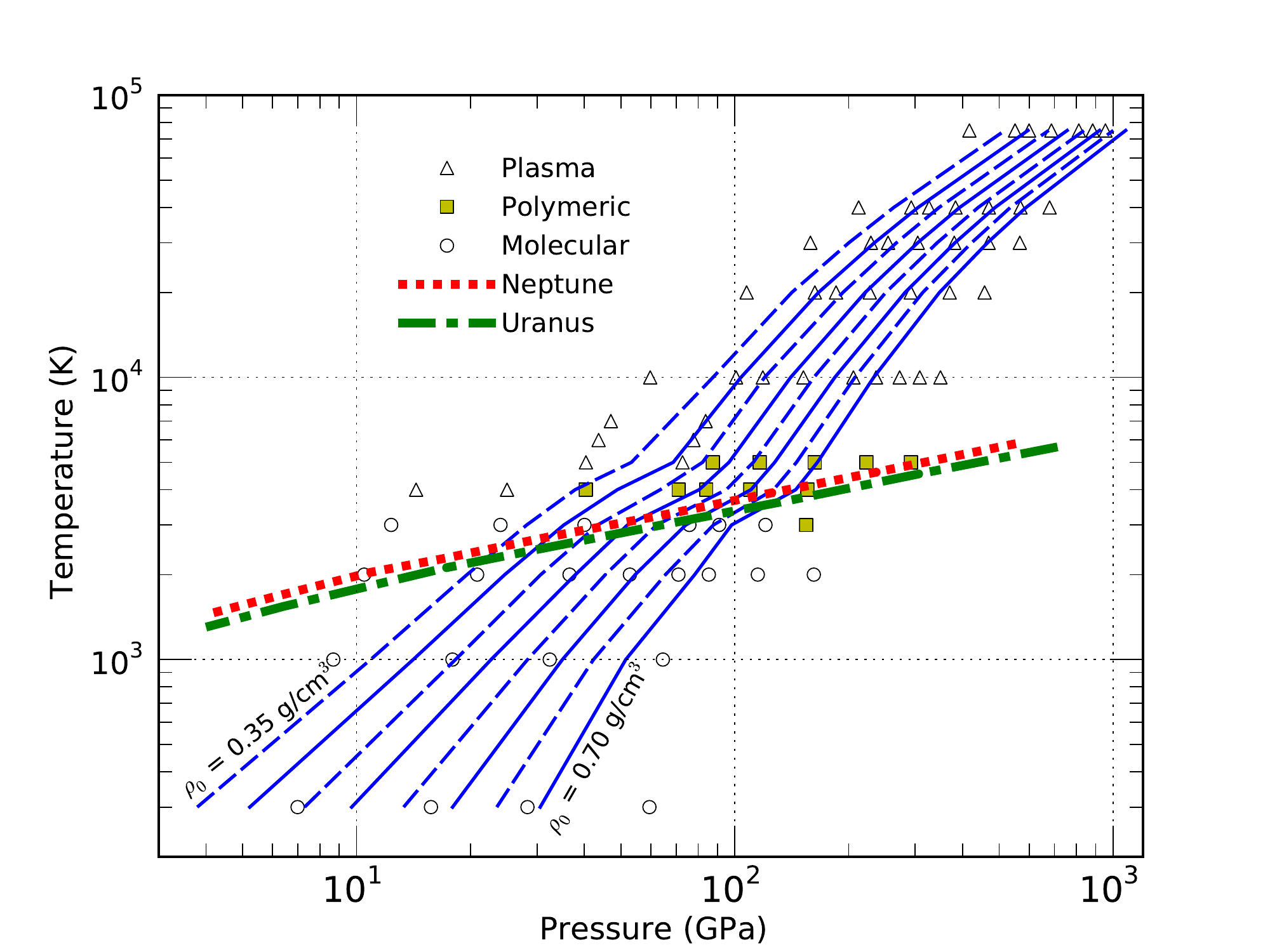}
\caption{Pressure as a function of temperature along the shock
  Hugoniot curves from Fig.~\ref{figure1} for different initial
  densities, $\rho_0$. Shown for reference are the Uranus and Neptune
  isentropes~\cite{redmer}.}
\label{figure2old}
\end{figure}

This cluster analysis allowed us to characterize all simulations in
Figs.~\ref{figure1} and \ref{figure2old} as either molecular,
polymeric, or as a plasma. It is important to note that the
classification refers to the state reached in our picosecond-timescale
DFT-MD simulations and, in case where we find the CH$_4$ molecules to
be stable, this may not necessarily be the thermodynamic ground state.
If all methane molecule remained stable in our simulation we described
the state as molecular, even though our cooled simulations did not
revert back to a molecular state. Previous studies (Ref.~\cite{spanu})
showed the polymeric state is thermodynamically favorable for
pressures above 4 GPa.

In Fig.~\ref{figure2old}, we plot temperature as a function of
pressure along each of the Hugoniot curves. There is a noticeable
change in the slope at approximately 4000--5000~K as the the Hugoniot
curves enter the polymeric regime and then the plasma state, but it is
not as pronounced as indicated in the temperature-density plot in
Fig.~\ref{figure1}.

We also included in Fig.~\ref{figure2old} the isentropes for Uranus
and Neptune that were approximately determined using DFT-MD
simulations of only H$_2$O~\cite{redmer}. These indicate that
precompressed shock experiments with initial densities from 0.35 to
0.70 g$\,$cm$^{-3}$ will be capable of reaching Uranian or Neptunian
interior conditions over a pressure range from 20 to 150 GPa. While
this pressure range can also be accessed with static diamond anvil cell
experiments, the corresponding temperature range of 2000--4000$\,$K
may be difficult to reach.

More importantly, Figure~\ref{figure2old} shows that the predicted
planetary isentropes pass through the polymeric regime. This implies
that any methane ice that was incorporated into the interiors of
Uranus and Neptune will not have remained there in molecular form.
Based on our simulations, we predict it to transform into a
polymeric state permitting a substantial amount of molecular hydrogen to
be expelled. The hydrogen gas may be released into the outer layer
that is rich in molecular hydrogen. 

It is also possible that more complex chemical reactions may take
place with a dense mixture of water, ammonia, and methance ice in the
interiors of Uranus and Neptune. It could also be difficult to
determine the state of chemical and thermodynamic equilibrium of the
ice mixture with a single DFT-MD simulation for a given
stochiometry because even very long simulations may remain in the
metastable state as we have seen here for methane.

The fact that hydrogen gas may be expelled from hot, dense
hydrocarbons may also imply that no superionic state of methane, in which the carbon atoms remain in their lattice positions
while the hydrogen atoms move through the crystal like a fluid, exists.
Superionic behavior has been predicted theoretically for hot, dense
water and ammonia~\cite{cavazzoni} but no predictions have been made
for methane despite the fact that the ratio of ionic radii of both
species are comparable to those in H$_2$O and NH$_3$. Since a
superionic state requires high pressure and temperatures, methane may
never transform into such a state because hydrogen gas may be released
instead. The remaining structure may be too dense for hydrogen
diffusion to occur.

	 
\begin{figure}[htbl]
\includegraphics[width=0.47\textwidth]{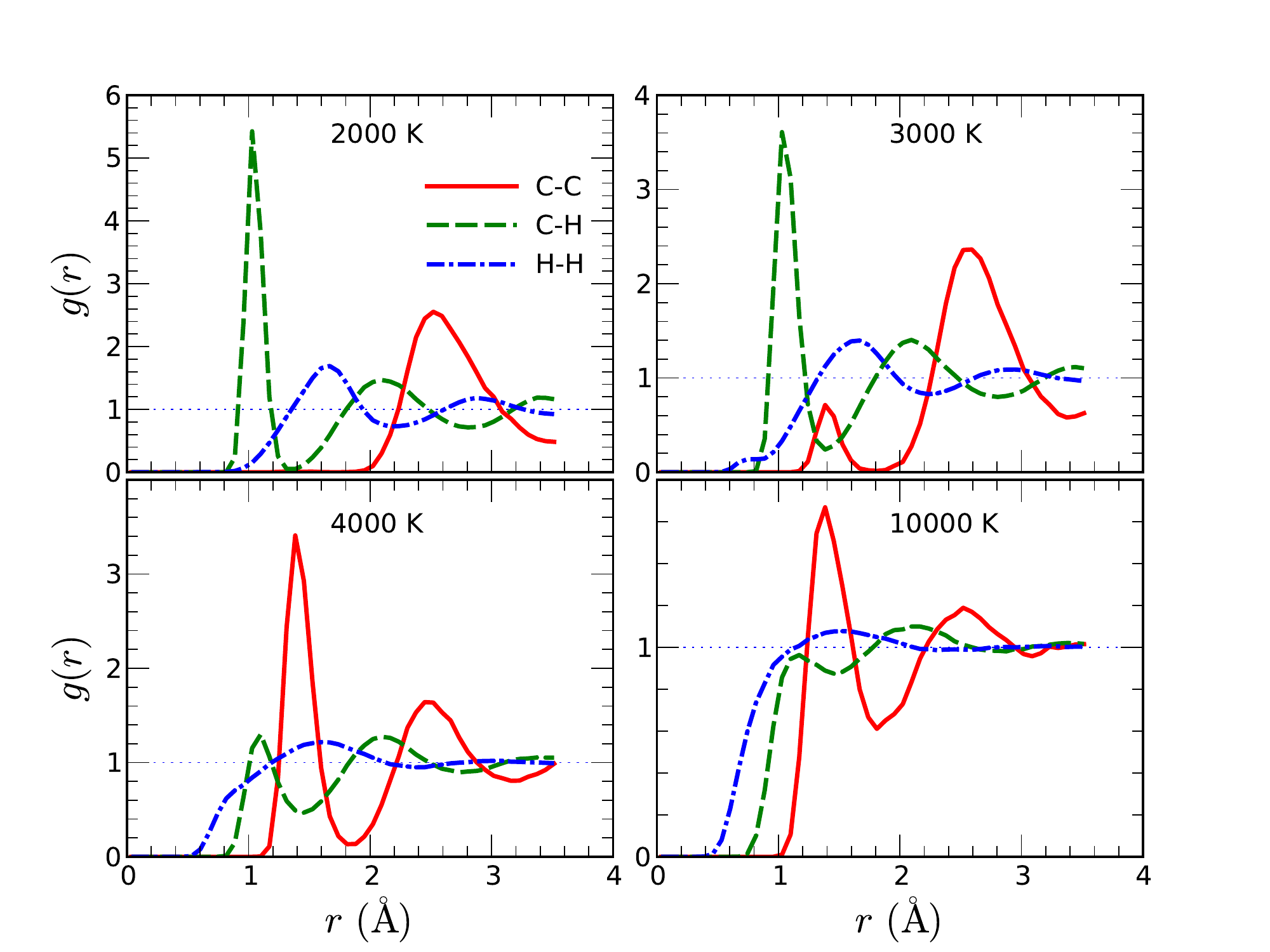}
\caption{Pair correlation functions, $g(r)$, between different nuclei
  as a function of temperature at a density of 2.0 g$\,$cm$^{-3}$.
  The intramolecular C-H and H-H peaks at 1.05 and
  1.7$\,$\AA~disappear as CH$_4$ molecules dissociate with increasing
  temperature, while the intensity of the C-C peak at 1.4$\,$\AA~rises
  with increased polymerization and then drops again as a plasma state
  is reached at 10$\,$000$\,$K.}
\label{figure4}
\end{figure}

The structure of the fluid was further investigated by examination of
pair correlation functions as shown in Fig.~\ref{figure4}. All pair
correlation functions were computed at the density 2.0~g$\,$cm$^{-3}$.
At 2000 K, methane molecules clearly dominate the system. The large
peak at 1.05 \AA~corresponds to the C--H bond of methane. This peak is
followed by a minimum close to zero, which is indicative of stable
molecules. When the temperature is raised to 3000 K, a new C--C peak
begins to emerge at 1.4 \AA, corresponding to the carbon bonds in
longer hydrocarbon chains. A corresponding decrease in the C--H peak,
and the appearance of a H--H peak of molecular hydrogen, is also seen.
At 4000 K, these processes become more apparent, as the C--C peak at
1.4 \AA~intensified and the C--H peak diminished. The flattening of
the H--H curve is a consequence of the dissociation of the methane
molecules. Lastly, at 10$\,$000 K the C--C peak, while reduced, is
still evident. However, at this temperature all molecules are very
short lived and unstable, and the broadening of C--H and H--H peaks
illustrate this fact.

\begin{figure}[htbl]
\includegraphics[width=0.47\textwidth]{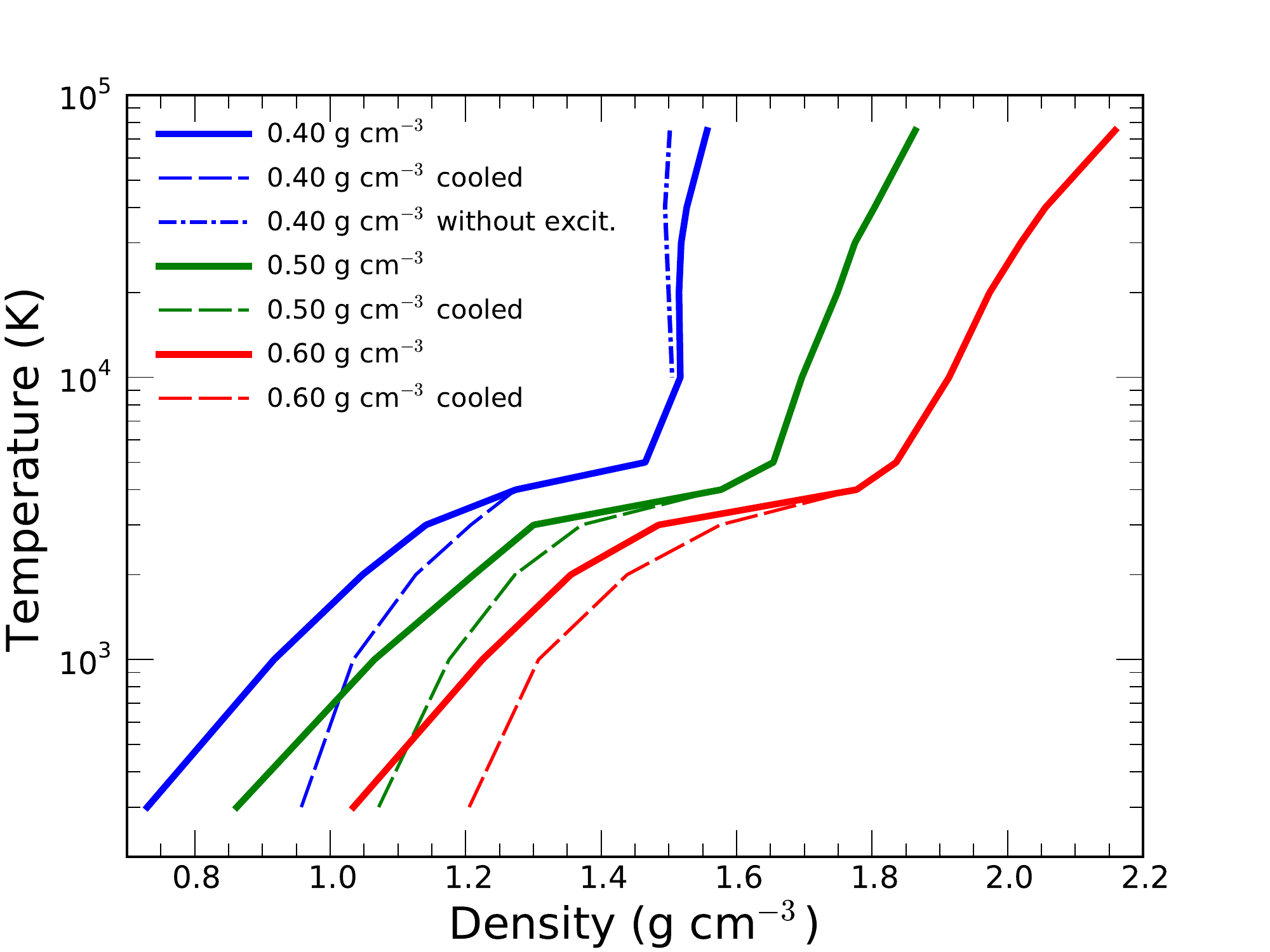}
\caption{Subset of Hugoniot curves from Fig.~\ref{figure1} for three initial
  densities, $\rho_0$, specified in the legend. The dashed lines
  correspond to polymeric samples that have been cooled from
  simulations at 4000~K. The dash-dotted line refers to simulations
  without contributions from excited electronic states.}
\label{figure5}
\end{figure}

We predicted the shock Hugoniot curves for the polymeric state by
cooling the simulations from 4000 to 300~K at six different densities
from 0.8 to 2.0~g$\,$cm$^{-3}$. Starting from the geometries and
velocities at 4000~K, we lowered the temperature in intervals of
1000~K by rescaling the velocities and performing MD simulations for
2~ps at each step. The structure of hydrocarbons, that were initially
present, change very little during the cooling simulations because the
temperature dropped below the activation barrier for such reactions.
As expected, all simulations eventually froze. For
$\rho=1.2\,$g$\,$cm$^{-3}$, it attained a composition of C$_9$H$_{20}$
+ C$_6$H$_{14}$ + 2C$_2$H$_6$ + 8CH$_4$ + 15H$_2$. These finding resemble
results of the shock experiment by Hirai {\emph et al.}~\cite{hirai},
who observed that the hydrocarbons produced did not revert back to methane
when the samples were cooled.

Figure~\ref{figure5} shows the Hugoniot curves for a cooled simulation
corresponding to three different initial densities.  At 4000~K, they
converge to our original Hugoniot curves but deviations become
increasingly pronounced with decreasing temperature, culminating in a
density increase of, respectively, 0.23 (31\%), 0.21 (24\%), and 0.17
g$\,$cm$^{-3}$ (24\%) for the initial densities of $\rho_0$ = 0.4,
0.5, and 0.6~g$\,$cm$^{-3}$ at 300~K. Because we kept the initial
parameters, $E_0$ and $V_0$, of methane in Eq.~\ref{eq1}, the
deviations between the original and the polymeric Hugoniot curves
persist in the limit of zero temperature. We predict the findings of a
future shock experiment on methane to fall in between both sets of
curves. For low shock velocities, one would expect the results to
track our original Hugoniot curves for stable molecules. For higher
shock speeds, the sample in the experiment may convert to a polymeric
state at lower temperatures than we see in our simulations. In this
case, we predict a significant increase in density to occur that
shifts the measured Hugoniot curve from the original towards the polymeric
Hugoniot curve. Figure~\ref{TP5} suggests that this may be
accompanied by a drop in pressure and a modest change in optical
properties. 

In shock experiments, one may observe the transformation into a
polymeric state at lower temperatures than we predict with our
simulations because experiments last for nanoseconds and samples are
much larger. However, also experiments may also not reach chemical
equilibrium at lower temperatures because the activitation energies
for different chemical reactions may be higher than the available
kinetic energy that could trigger such reaction.

The fact that the molecular and the polymeric Hugoniot curve diverge
below 4000~K in Fig.~\ref{figure5} also has implications for decaying
shock experiments~\cite{eggert10}. This new experimental technique
allows one to map out a whole segment of the Hugoniot curve with a
single shock wave experiment. At the beginning of the shock
propagation, the material in compressed to a state of high pressure
and temperature on the Hugoniot curve. As the shock wave continues to
propagate, the particle velocity is gradually reduced and new material
is compressed to a lower pressure and temperature. The shock velocity
is continuously monitored with an interferometer so that a whole
segment of Hugoniot curve can be measured in one shock experiment. If
a decaying shock acrosses the polymeric regime of methane, we predict
the following to occur. As the shock enters the polymeric regime, new
material at the shock front would still be compressed as in a usual
shock experiments and may yield a state on the molecular or on the
polymeric Hugoniot, as we discussed. However, as material behind the
shock front expands from a high-temperature state, it will change from
a plasma into the polymeric state rather than converting back to
molecular methane, as we have observed in our cooled simulations in
Fig.~\ref{TP5}. This may lead to a structured shock wave with various
parts in different thermodynamic states. So the polymerization of
methane may provide a mechanism for introducing a shock velocity
reversal that were observed in ~\cite{Spaulding2012} and discussed
in~\cite{Militzer2012}.

Finally, we investigated the effects that thermal electronic excitation
have on the shape of the Hugoniot curves. While electronic excitations
were included in all simulations discussed so far, we performed some
DFT-MD simulations where the electrons were kept in the ground state.
This resulted into drastic reductions of the computed pressures and
internal energies but both effects nearly cancelled~\cite{Mi06} when
the Hugoniot curve was calculated. So Figure~\ref{figure5} shows only
a modest shift towards lower densities for temperatures over
10$\,$000$\,$K.

\section{Conclusions}

We identified and characterized molecular, polymeric, and plasma
states in the phase diagram of methane by analyzing the of the liquid
in DFT-MD simulations that we performed in a temperature-density range
from 300 to 75$\,$000~K and 0.8 to 2.5~g$\,$cm$^{-3}$. We presented
shock Hugoniot curves for initial densities from 0.35 to 0.70
g$\,$cm$^{-3}$. We predict a drastic increase in density along the
Hugoniot curves as the sample transforms into a polymeric state.  This
transformation is accompanied by an increase in reflectivity and
electrical conductivity because the polymeric state is metallic.

This state is also prone to dehydrogenation reactions, which has
implications for the interiors of Uranus and Neptune. These planets' isentropes
intersect the temperature-pressure conditions of the polymeric regime
and thus molecular hydrogen may be released from ice layers of both
planets. This could lead to more compact cores in Uranus and Neptune,
in contrast to recent predictions of core erosion for
Jupiter and Saturn~\cite{WilsonMilitzer2012,WilsonMilitzer2012b}.

We also argued that the dehydrogenation reactions prevent methane from
assuming a superionic state at high pressure and temperature that was
predicted for water and ammonia~\cite{cavazzoni}. However more
theoretical and experimental work on mixture of planetary ices will be
required to provide much-needed constraints for the interiors of ice
giant planets~\cite{NewHelledPaper}.


{\bf Acknowledgements:} This work was supported by NASA and NSF.
Computational resources were supplied in part by NCCS and TAC.

\appendix

\section*{The following table is to be published as online supplementary information}
\begin{table}
\centering
{\footnotesize
\begin{ruledtabular}
\begin{tabular}{r c l }
$T$ (K)         & $P$ (GPa) & $E$ (eV/CH$_4$)\\
\hline
&$\rho$=  0.600 g$\,$cm$^{-3}$\\ 
   300 & 1.957(10)  &  $-$0.173451(13)\\
\hline
&$\rho$=  0.800 g$\,$cm$^{-3}$\\ 
   300 & 6.984(18)  &  $-$0.172014(10)\\
  1000 & 8.68(4)  &  $-$0.16580(5)\\
  2000 & 10.47(3)  &  $-$0.15674(3)\\
  3000 & 12.35(9)  &  $-$0.14831(8)\\
  4000 & 14.37(11)  &  $-$0.1354(3)\\
\hline
&$\rho$=  1.000 g$\,$cm$^{-3}$\\ 
   300 & 15.73(4)  &  $-$0.169205(11)\\
  1000 & 17.93(4)  &  $-$0.16274(2)\\
  2000 & 20.84(11)  &  $-$0.15374(6)\\
  3000 & 24.02(10)  &  $-$0.14456(8)\\
  4000 & 25.00(13)  &  $-$0.1321(2)\\
\hline
&$\rho$=  1.201 g$\,$cm$^{-3}$\\ 
   300 & 28.309(19)  &  $-$0.165345(9)\\
  1000 & 32.42(5)  &  $-$0.15855(2)\\
  2000 & 36.57(12)  &  $-$0.14920(4)\\
  3000 & 40.05(11)  &  $-$0.13961(8)\\
  4000 & 40.4(3)  &  $-$0.1237(4)\\
  5000 & 40.41(14)  &  $-$0.1067(2)\\
  6000 & 43.65(16)  &  $-$0.0960(2)\\
  7000 & 47.0(3)  &  $-$0.08730(11)\\
 10000 & 59.76(9)  &  $-$0.06481(12)\\
 20000 & 107.47(20)  &  $\;\;\;$0.00630(11)\\
 30000 & 158.44(16)  &  $\;\;\;$0.08227(10)\\
 40000 & 212.5(3)  &  $\;\;\;$0.16452(12)\\
 75000 & 417.0(3)  &  $\;\;\;$0.49150(13)\\
\hline
&$\rho$=  1.353 g$\,$cm$^{-3}$\\ 
  2000 & 52.81(10)  &  $-$0.14496(5)\\
\hline
&$\rho$=  1.498 g$\,$cm$^{-3}$\\ 
   300 & 59.48(3)  &  $-$0.157294(10)\\
  1000 & 64.55(8)  &  $-$0.15038(3)\\
  2000 & 70.97(16)  &  $-$0.14057(5)\\
  3000 & 75.87(17)  &  $-$0.13066(6)\\
  4000 & 71.09(15)  &  $-$0.1126(3)\\
  5000 & 72.76(19)  &  $-$0.0984(2)\\
  6000 & 77.7(4)  &  $-$0.08951(14)\\
  7000 & 83.7(2)  &  $-$0.0815(2)\\
 10000 & 100.79(14)  &  $-$0.05941(8)\\
 20000 & 162.79(20)  &  $\;\;\;$0.01036(11)\\
 30000 & 229.0(6)  &  $\;\;\;$0.0870(11)\\
 40000 & 292.7(1.7)  &  $\;\;\;$0.1658(10)\\
 75000 & 550.8(4)  &  $\;\;\;$0.48502(13)\\
\hline
&$\rho$=  1.600 g$\,$cm$^{-3}$\\ 
  2000 & 85.39(16)  &  $-$0.13698(5)\\
  3000 & 90.95(14)  &  $-$0.12687(11)\\
  4000 & 84.02(19)  &  $-$0.1086(2)\\
  5000 & 87.5(2)  &  $-$0.09565(18)\\
 10000 & 118.6(2)  &  $-$0.05693(9)\\
 20000 & 185.2(4)  &  $\;\;\;$0.0129(4)\\
 30000 & 254.2(4)  &  $\;\;\;$0.08663(10)\\
 40000 & 326.4(3)  &  $\;\;\;$0.16661(18)\\
 75000 & 599.8(5)  &  $\;\;\;$0.4841(2)\\
\end{tabular}
\end{ruledtabular}}
\end{table}
\begin{table}
\centering
{\footnotesize
\begin{ruledtabular}
\begin{tabular}{r c l }
$T$ (K)         & $P$ (GPa) & $E$ (eV/CH$_4$)\\
\hline
&$\rho$=  1.775 g$\,$cm$^{-3}$\\ 
  2000 & 115.04(19)  &  $-$0.13039(4)\\
  3000 & 120.6(3)  &  $-$0.11986(11)\\
  4000 & 110.0(4)  &  $-$0.1018(2)\\
  5000 & 116.36(13)  &  $-$0.09025(13)\\
 10000 & 151.85(18)  &  $-$0.05220(9)\\
 20000 & 227.5(3)  &  $\;\;\;$0.01736(12)\\
 30000 & 304.6(5)  &  $\;\;\;$0.0917(6)\\
 40000 & 383.3(9)  &  $\;\;\;$0.1700(2)\\
 75000 & 687.1(8)  &  $\;\;\;$0.4836(3)\\
\hline
&$\rho$=  2.010 g$\,$cm$^{-3}$\\ 
  2000 & 161.77(17)  &  $-$0.12039(6)\\
  3000 & 154.4(4)  &  $-$0.10612(16)\\
  4000 & 155.5(2)  &  $-$0.09276(11)\\
  5000 & 162.68(13)  &  $-$0.08206(10)\\
 10000 & 205.9(3)  &  $-$0.04411(10)\\
 20000 & 292.0(4)  &  $\;\;\;$0.0250(2)\\
 30000 & 380.4(3)  &  $\;\;\;$0.09839(18)\\
 40000 & 469.6(6)  &  $\;\;\;$0.1763(3)\\
 75000 & 811.9(6)  &  $\;\;\;$0.4857(2)\\
\hline
&$\rho$=  2.129 g$\,$cm$^{-3}$\\ 
 10000 & 236.22(19)  &  $-$0.03969(9)\\
\hline
&$\rho$=  2.257 g$\,$cm$^{-3}$\\ 
  5000 & 222.7(3)  &  $-$0.07214(11)\\
 10000 & 272.8(5)  &  $-$0.03438(14)\\
 20000 & 369.9(4)  &  $\;\;\;$0.03459(13)\\
 30000 & 468.9(6)  &  $\;\;\;$0.10728(18)\\
 40000 & 569.2(6)  &  $\;\;\;$0.1843(2)\\
 75000 & 954.7(6)  &  $\;\;\;$0.49147(16)\\
\hline
&$\rho$=  2.376 g$\,$cm$^{-3}$\\ 
 10000 & 308.3(4)  &  $-$0.02945(12)\\
\hline
&$\rho$=  2.502 g$\,$cm$^{-3}$\\ 
  5000 & 292.4(5)  &  $-$0.06108(13)\\
 10000 & 349.7(5)  &  $-$0.02343(16)\\
 20000 & 457.5(2)  &  $\;\;\;$0.04569(10)\\
 30000 & 566.9(9)  &  $\;\;\;$0.1180(2)\\
 40000 & 679.6(5)  &  $\;\;\;$0.19499(14)\\
 75000 & 883.4(7)  &  $\;\;\;$0.32519(17)\\
\end{tabular}
\end{ruledtabular}} 
\caption{Pressure and internal energy from our 
  DFT-MD simulations at various densities and temperatures. 
  The 1 $\sigma$ error bars are given in brackets.
}
\label{tab1}
\end{table}

\end{document}